\documentclass[prl,aps,showpacs,twocolumn]{revtex4}
\usepackage{graphics,bm}
\usepackage{amssymb}
\usepackage{epsfig}
\usepackage{epsf}
\usepackage[usenames]{color}

\begin{document}

\title{Current-induced torques in continuous antiferromagnetic textures}

\author{A.C. Swaving}

\author{R.A. Duine}

\affiliation{Institute for Theoretical Physics, Utrecht
University, Leuvenlaan 4, 3584 CE Utrecht, The Netherlands}
\date{\today}

\begin{abstract}

We study the influence of an electric current on a continuous
non-collinear antiferromagnetic texture. Despite the lack of a net
magnetic moment we find that the exchange interaction between
conduction electrons and local magnetization generally results in
current-induced torques that are similar in phenomenology to spin
transfer torques in ferromagnets. We present the generalization of
the non-linear sigma model equation of motion for the N\'{e}\`{e}l
vector that includes these current-induced torques, and briefly
discuss the resulting current-induced antiferromagnetic domain
wall motion and spin-wave Doppler shift. We give an interpretation
of our results using a unifying picture of current-induced torques
in ferromagnets and antiferromagnets in which they are viewed as
due to the current-induced spin polarization resulting from an
effective spin-orbit coupling.

\end{abstract}

\maketitle

\def\bx{{\bf x}}
\def\bk{{\bf k}}
\def\half{\frac{1}{2}}
\def\args{(\bx,t)}

\noindent {\it Introduction} --- Magnetoresistive phenomena in
conducting ferromagnets, resulting from the interplay between
spins of conduction electrons and the magnetization, are well
known. They play a major role in new technologies and are key to
the recently awarded Nobel prize for Giant
Magnetoresistance\cite{grunberg1986,fert1988}. Recently, the
effect of a spin current on magnetization dynamics --- called spin
transfer --- has been investigated in detail
\cite{berger1996,slonczewski1996,tsoi1998,myers1999}. Although
applications, mainly to memory storage technology, are an
important driving factor behind this research, spin transfer is
also physically interesting in its own right. It can be understood
as follows.

Consider a conducting ferromagnet far below its critical
temperature such that it is described by a unit vector
$\vec{\Omega}(\vec{x},t)$ in the direction of magnetization. Its
dynamics in the presence of an effective field,
$\vec{H}_{\mathrm{eff}}$, is determined by
\begin{equation}
\label{eq:magdynam} \frac{\partial\vec{\Omega}}{\partial
t}=\vec{\Omega}\times
\vec{H}_{\mathrm{eff}}-J\vec{\Omega}\times\vec{\nabla}^{2}\vec{\Omega},
\end{equation}
where we have ignored magnetization relaxation and the effective
field contains in first instance contributions from the anisotropy
of the system and the external field. Furthermore, $J$ is the
exchange constant favoring alignment of neighboring spins. In the
presence of conduction electrons the effective field contains an
additional contribution due to the $s-d$ exchange coupling of the
magnetization with the spin density $\vec{s}$ of the conduction
electrons given by $-\int
d\vec{x}\Delta\vec{s}\cdot\vec{\Omega}/a^{3}$, with $\Delta$ the
exchange splitting and $a$ the lattice constant. (Although the
$s-d$ model is convenient to illustrate the physics involved the
conclusions drawn are qualitatively valid for other models as
well.) In equilibrium, i.e., without a current, this spin density
will be in the plane of the magnetization. However, when a current
is applied, the spin density acquires a component
$\left<\vec{s}_{\perp}\right>$ perpendicular to the magnetization
direction $\vec{\Omega}$. It is this component of the spin density
that leads to current-driven dynamics of the magnetization
\cite{Nunez2006} by contributing to the effective field as
\begin{equation}
\label{eq:efffield}
\vec{H}_{\mathrm{eff}}\Big|_{\mathrm{current}}=\frac{\Delta a^{3}
}{\hbar}\left<\vec{s}_{\perp}\right>.
\end{equation}
To lowest order in the current and the gradient of the
magnetization, and using an adiabatic assumption based on the fact
that the electron dynamics is fast compared to the time variation
of the magnetization, the out-of-plane spin density is
\begin{equation}
\label{eq:spindensity}
\left<\vec{s}_{\perp}\right>=\frac{\hbar}{\Delta a^{3}
}\vec{\Omega}\times\left(\vec{\upsilon}_{s}\cdot\vec{\nabla}\right)\vec{\Omega},
\end{equation}
where the velocity $\vec{\upsilon}_{s}$, defined via the above
equation, is proportional to the electric current. Inserting the
above contribution to $\vec{H}_{\mathrm{eff}}$ in
Eq.(\ref{eq:magdynam}) we find that in the steady-state transport
situation the contribution of the current-induced torques to the
equation of motion for the magnetization direction is given by
\begin{equation}
\label{eq:FMeqm} \frac{\partial\vec{\Omega}(\vec{x},t)}{\partial
t}\Big|_{\mathrm{current}}=-\left(\vec{\upsilon}_{s}\cdot\vec{\nabla}\right)\vec{\Omega}(\vec{x},t).
\end{equation}
The current-dependent velocity is in the absence of spin-orbit
coupling and spin-flip scattering fixed by spin conservation, as
follows. Integrating Eq.(\ref{eq:FMeqm}) over the length of the
ferromagnet in the direction $\hat r$ of the current leads to the
total change in angular momentum $\int\! dr\,\partial
\vec{\Omega}(\vec{x},t)/\partial
t\Big|_{\mathrm{current}}=\vec{\upsilon}_{s}\cdot \hat r
\left[\left.
\vec{\Omega}\right|_{\mathrm{in}}-\left.\vec{\Omega}\right|_{\mathrm{out}}\right]$,
where the current flows from in to out. By spin conservation, this
change in angular momentum is proportional to the change in the
spin current $\vec{J}_{s}^{\alpha}$ after it has passed through
the ferromagnet, i.e., $\int\! dr\,\partial
\Omega^\alpha(\vec{x},t)/\partial
t\Big|_{\mathrm{current}}=\vec{J}^{\alpha}_{s,\mathrm{in}}-\vec{J}^{\alpha}_{s,\mathrm{out}}$.
Using that $\vec{J}_{s}^{\alpha}=P\vec{J_{c}}\Omega^{\alpha}$,
with $P$ the polarization of the charge current $\vec{J}_c$ in the
ferromagnet, we find that $\vec{\upsilon}_{s}\propto
P\vec{J}_{c}$. This argument shows that the current-induced torque
in Eq.~(\ref{eq:FMeqm}) results from transfer of angular momentum
from conduction electrons to magnetization, and is hence called a
spin transfer torque.

Well-known examples of dynamics resulting from these spin transfer
torques involve spin waves and domain
walls\cite{bazaliy1998,Fernandez2004,tatara2004,thiaville2005,grollier2003,tsoi2003,yamaguchi2004,klaui2005,beach2006,hayashi2007,yamanouchi2004,Vlaminck2008}.
For spin waves a Doppler shift in the dispersion relation is found
once a current is applied \cite{bazaliy1998,Fernandez2004}. This
shift is linearly proportional to $\vec{k}$ where the
proportionality constant is given by $\vec{\upsilon}_{s}$ and the
dispersion that follows from
Eqs.~(\ref{eq:magdynam}-\ref{eq:FMeqm})is then given by
$\hbar\omega\!\!=\!\!J\vec{k}^{2}\!\!+\!\vec{\upsilon}_{s}\cdot
\vec{k}$, where we ignored anisotropy and external fields.
Recently experiments have successfully measured such
current-induced spin-wave Doppler shifts\cite{Vlaminck2008}.
Current-induced domain wall motion
\cite{tatara2004,thiaville2005,grollier2003,tsoi2003,yamaguchi2004,klaui2005,beach2006,hayashi2007,yamanouchi2004}
is understood as follows. In the absence of any pinning and
damping, $\vec{\Omega}_{0}(\vec{x}-\vec{\upsilon}_{s}t)$ is a
solution to Eq.(\ref{eq:FMeqm}), where $\vec{\Omega}_{0}(\vec{x})$
is a time-independent solution of Eq.(\ref{eq:magdynam}). For the
case that $\vec{\Omega}_{0}(\vec{x})$ corresponds to a domain wall
texture this means that the velocity of the domain wall is just
$\vec{\upsilon}_{s}$, an estimate which turns out to be reasonable
even when pinning and damping are present
\cite{tatara2004,thiaville2005}. These results are also understood
by realizing that Eq.(\ref{eq:FMeqm}) shows that it is possible to
get the equation of motion of $\vec{\Omega}$ with an applied
current by substituting $\frac{\partial}{\partial
t}\rightarrow\frac{\partial}{\partial
t}+\vec{\upsilon}_{s}\cdot\vec{\nabla}$ into the equation of
motion without current.

In antiferromagnets adjacent lattices sites have opposing magnetic
moments and thus form two sublattices with opposite magnetization
leaving no net magnetization. They are characterized by a
N\'{e}\`{e}l vector $\vec{n}_{j}=(-1)^{j}\vec{\Omega}_{j}$ where
$j$ labels the lattice sites. Its equation of motion is a
non-linear sigma model in the presence of an effective field,
$\vec{H}_{\mathrm{eff}}$, and is given by
\begin{equation}
\label{eq:AFMnlsm} \frac{\partial^{2}\vec{n}}{\partial
t^{2}}-\frac{\partial}{\partial
t}\left(\vec{n}\times\vec{H}_{\mathrm{eff}}\right)-c^{2}\nabla^{2}\vec{n}=0,
\end{equation}
to lowest order in $\vec{H}_{\mathrm{eff}}$ and where $c$ is the
spin-wave velocity. In this Letter we show that in the presence of
an electric current in an antiferromagnetic conductor there is, to
lowest order in N\'{e}\`{e}l-vector gradient and the current, an
out-of-plane spin density given by
\begin{equation}
\label{eq:afmspinden}
\left<\vec{s}_{\perp}\right>=\frac{\hbar}{\Delta a^{3}
}\vec{n}\times\left(\vec{\upsilon}\cdot\vec{\nabla}\right)\vec{n},
\end{equation}
where $\vec{\upsilon}$, defined with the above equation, is
proportional to the current and the antiferromagnetic equivalent
of the velocity $\vec{\upsilon}_{s}$ that was introduced in
describing current-induced torques in ferromagnets. Although the
form of the spin density is the same as for a ferromagnet, the
velocity $\vec{\upsilon}$ cannot be determined from macroscopic
spin conservation arguments, as in the case of the ferromagnet,
but instead needs to be determined by a microscopic calculation
(such a calculation is discussed in detail below). Using the
equation of motion, Eq.(\ref{eq:AFMnlsm}), and the spin density in
Eq.(\ref{eq:afmspinden}) as a contribution to
$\vec{H}_{\mathrm{eff}}$ gives the equation of motion
\begin{equation}
\label{eq:AFMeqm} \frac{\partial^{2}\vec{n}}{\partial
t^{2}}+\frac{\partial}{\partial
t}(\vec{\upsilon}\cdot\vec{\nabla})\vec{n}-c^{2}\nabla^{2}\vec{n}=0~,
\end{equation}
where we ignored anisotropy. The antiferromagnetic spin wave
dispersion resulting from this equation is, again ignoring
anisotropy and external fields, given by $\hbar\omega=ck +
\vec{\upsilon} \cdot \vec{k}/2$ to linear order in current. This
shows that the dispersion has a similar Doppler shift as the
ferromagnetic case \cite{bazaliy1998,Fernandez2004}. Similarly
like in the ferromagnetic case, the equation of motion in
Eq.~(\ref{eq:AFMeqm}) allows for co-moving solutions $n_0 (\vec{x}
- \vec{\upsilon} t/2)$ (to first order in $\vec{\upsilon}$), from
which we conclude that antiferromagnetic domain walls
\cite{Papanicolaou1995} move with velocity $\vec{\upsilon}/2$
proportional to the current in the absence of pinning and damping.
Note that the above equation of motion is obtained by replacing
$\frac{\partial}{\partial t}\rightarrow\frac{\partial}{\partial
t}+\vec{\upsilon}\cdot\vec{\nabla}/2$ in Eq.~(\ref{eq:AFMnlsm})
and keeping terms to first order in $\vec{\upsilon}$.

So far we have suggested a form for the spin density of the
conduction electrons in an antiferromagnetic metal in the presence
of current, i.e.,
$\left<\vec{s}_{\perp}\right>\propto\vec{n}\times\left(\vec{\upsilon}\cdot\vec{\nabla}\right)\vec{n}$.
In the remainder of this Letter we prove this result and show that
$\vec{\upsilon}$ is generally not zero and can be of the same
order of magnitude as $\vec{v_{s}}$, which is of the order of
$1-100$ m/s for a typical ferromagnetic alloy like permalloy. We
calculate $\vec{v}$ for a toy-model antiferromagnetic metal.
Further, we explain our results using a unifying picture of
current-induced torques in ferromagnetic and antiferromagnetic
textures, that is built on the notion of effective spin-orbit
coupling induced by a non-collinear magnetic texture.

Previous work on current-induced torques in antiferromagnetic
metals considered mainly single-domain layered structures
\cite{nunez2006,wei2007,urazhdin2007,tang2007,herranz2009} and the
situation of an antiferromagnetic domain-wall was considered from
an ab initio point-of-view \cite{xu2008}. In this Letter we derive
the general phenomenology of current-induced torques in spatially
smooth and slowly-varying antiferromagnetic textures.

\begin{figure}[h]
\scalebox{0.5}{\includegraphics[0,0][300,46]{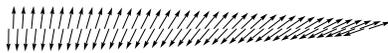}}
\caption[Picture of spin-wave]{Illustration of a smooth
antiferromagnetic magnetization texture.} \label{figure 1}
\end{figure}

\begin{figure}[h]
\begin{center}
\scalebox{0.5}{\includegraphics[0,0][459,296]{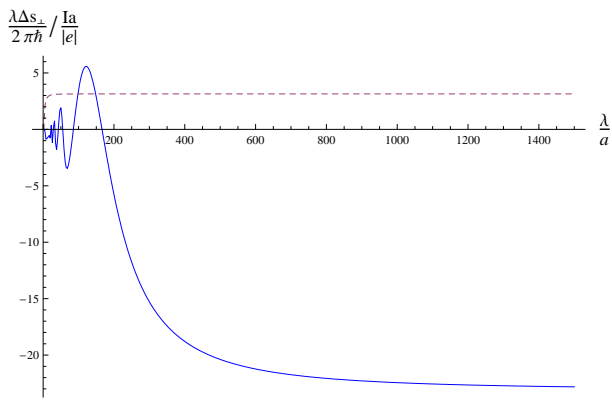}}
\end{center}
\caption[Plot of velocity vs Lambda]{Plot of the spin density
versus the texture wavelength, $ \lambda$. The solid line
corresponds to the antiferromagnet and the dashed line represents
the ferromagnet.} \label{figure 2}
\end{figure}

\noindent {\it Toy-Model Antiferromagnetic Metal} --- To confirm
the conjecture
$\left<\vec{s}_{\perp}\right>\propto\vec{n}\times(\vec{\upsilon}
\cdot  \vec{\nabla})\vec{n}$ and determine the current-dependent
velocity $\vec{\upsilon}$ we consider a toy model of a conducting
antiferromagnet within the Green's function formulation of
Landauer-B\"{u}ttiker transport theory\cite{dattabook}. The
tight-binding hamiltonian used in this theory is given by
$H=H_{S}+H_{L}+H_{I}$ where
\begin{equation}
\label{eq:TBHamitonian}
H_{S}=-t\sum_{<j,j'>;\sigma}\psi_{j,\sigma}^{\dag}\psi_{j',\sigma}
-\sum_{j;\sigma,\sigma'}\psi_{j,\sigma}^{\dag}\left[\frac{\Delta}{2}\vec{\Omega_{j}}\cdot\vec{\tau}_{\sigma,\sigma'}\right]\psi_{j,\sigma'},
\end{equation}
is the system Hamiltonian with $t$ the nearest-neighbor hopping
amplitude and $\Delta$ the exchange energy, and $\psi_{j,\sigma}$
are $\psi_{j,\sigma}^\dagger$ are the electron annihilation and
creation operators. $H_{L}$ and $H_{I}$ are the Hamiltonians for
the leads and for the coupling between the leads and the system
respectively and are similar but with $\Delta=0$. To realize a
transport current $I$ these leads have a chemical potential
difference of $e|V|$. The magnetization texture is set to
$\vec{\Omega_{j}}=((-1)^{j}\sin(\frac{2\pi
aj}{\lambda}),0,(-1)^{j}\cos(\frac{2\pi aj}{\lambda}))$ where
$\lambda$ is the wavelength of the antiferromagnetic texture. See
Fig.~\ref{figure 1} for an illustration.
This magnetization texture is in the $x-z$ plane and in
equilibrium, without current, the system only has nonzero spin
densities
$\vec{s}=\left<\psi_{\sigma}^{\dag}\vec{\tau}_{\sigma,\sigma'}\psi_{\sigma'}\right>$
in this plane. For nonzero voltage we find a spin density
$\vec{s}_{\perp}$ in the $y$-direction that is constant in
position, in agreement with Eq.~(\ref{eq:afmspinden}). According
to Eq.~(\ref{eq:afmspinden}) the velocity,
\begin{equation}
\label{eq:velocity}
\upsilon=\frac{\lambda\Delta
\left<s_{\perp}\right>}{2\pi\hbar},
\end{equation}
should be independent of $\lambda$ in the long-wavelength limit.
Fig.~\ref{figure 2} confirms this. The fluctuations for small
$\lambda$ in Fig.~\ref{figure 2} are due to finite size effects.
The ferromagnetic result is also shown in this figure. Note that
the long-wavelength limit is reached for smaller $\lambda$ in the
ferromagnetic case, compared to the antiferromagnetic one. This is
understood as the unit cell of the antiferromagnetic system is
effectively doubled because of the opposing magnetization on
neighboring sites.

\

\begin{figure}[h]
\begin{center}
\scalebox{0.5}{\includegraphics[0,0][459,297]{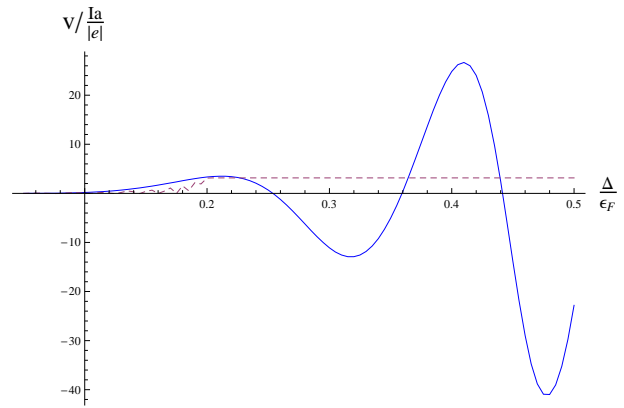}}
\end{center}
\caption[Plot of normalized velocity vs Delta]{Plot of the
velocity, $\upsilon$, versus the ratio of the exchange constant
and Fermi energy, $ \frac{\Delta}{\epsilon_{F}}$. The solid line
represents the antiferromagnet and the dashed line represents the
ferromagnet.} \label{figure 3}
\end{figure}

Now that we have established that $\vec{\upsilon}$ approaches a
constant in the long-wavelength limit, we take its long-wavelength
limiting value as its definition and study its dependence on
$\Delta$. The result is shown in Fig.~\ref{figure 3}. It is clear
from this figure that $\vec{\upsilon}$ depends strongly on the
value of the exchange constant. This dependence reflects the
strong dependence of the quasi-particle band structure on
$\Delta$. The ferromagnetic case is also shown for comparison.
Note that in the ferromagnetic case the velocity for increasing
$\Delta$ quickly obtains its value determined by spin
conservation. In the antiferromagnetic case the velocity is not
determined by a macroscopic conservation law. Note, however, that
the velocities $\vec{v}_s$ and $\vec{v}$, respectively
parametrizing current-induced torques in ferro and
antiferromagnets, are of the same order of magnitude.

\noindent {\it Effective Spin-Orbit Coupling} --- The above
numerical analysis confirms that when a current is applied to an
antiferromagnetic metal there is a nonzero out-of-plane spin
density that drives current-induced magnetization dynamics. We now
give an interpretation for this spin density in terms of a
current-induced spin polarization resulting from an effective
spin-orbit coupling. We start with a system of electrons (mass
$m$), moving in an antiferromagnetic texture $\vec{n} (\vec{x}_j)$
and scalar potential $V (\vec{x}_j)$, described by the Hamiltonian
\begin{equation}
\label{eq:generalHamiltonian}
H=\frac{p^{2}}{2m}+V(\vec{x_{j}})-\frac{\Delta}{2}(-1)^{j}\vec{n}(\vec{x_{j}})\cdot\vec{\tau},
\end{equation}
with $\vec{x}_{j}=ja$ the position of the $j$-th lattice site. We
align the spin quantization axis of the conduction electrons to
the local N\'{e}\`{e}l vector by applying a $SU(2)$ transformation
$\left|\psi\right>\rightarrow R\left|\psi\right>$ to the
wavefunction with $R^{-1}\vec{n}\cdot\vec{\tau}R=\tau^{z}$ that
therefore diagonalizes the spin part of the Hamiltonian and gives
the effective Hamiltonian
\begin{equation}
\label{eq:SU2Hamiltonian} H_{\rm
eff}=H_{0}-2iJ^{\mu}_{s,\alpha}a_{\alpha}^{\mu}(\vec{x}),
\end{equation}
to first order in the gradient of $\vec{n}$, where $H_{0}$ is the
Hamiltonian in Eq.~(\ref{eq:generalHamiltonian}) with $\vec{n} =
\hat z$. Here,
$J^{\mu}_{s,\alpha}=\frac{\hbar}{2}\tau^{\mu}v_{0,\alpha}$ is the
spin current with the velocity
$v_{0,\alpha}=\frac{1}{\hbar}\frac{\partial\epsilon_{\vec{k}}}{\partial
k_{\alpha}}$ with $\epsilon_{\vec{k}}$ the dispersion
corresponding to $H_{0}$ and the gauge fields
$a_{\alpha}^{\mu}(\vec{x})\cong
i\vec{n}\times\nabla_{\alpha}\vec{n}|_{\mu}$\cite{Shraiman1988}.
Substituting the latter into Eq.(8) leads to the effective
Hamiltonian
\begin{eqnarray}
\label{eq:SOHamiltonain} H_{\rm
eff}&=&H_{0}+\left(\vec{n}\times\left[\frac{\partial\epsilon_{\vec{k}}}{\partial
k_{\alpha}}\nabla_{\alpha}\right]\vec{n}\right)\cdot\vec{\tau}\nonumber\\
&\equiv&H_{0}-\vec{B}_{\mathrm{eff}}\cdot\vec{\tau},
\end{eqnarray}
where $\vec{B}_{\mathrm{eff}}$ is a momentum-dependent fictitious
magnetic field that can be viewed as an effective spin-orbit
coupling resulting from the non-collinear antiferromagnetic
texture. (Note that this effective spin-orbit coupling is
different from the result of Ref.~\cite{revaz2008} for a collinear
antiferromagnet.) Since the effective magnetic field is linearly
proportional to the velocity operator, when an electric field is
applied $B_{\mathrm{\mathrm{eff}}}$ is nonzero. The resulting
Zeeman splitting of the electron spins due to $B_{\mathrm{eff}}$
results in a current-induced spin polarization that is aligned
with the effective magnetic field and, since $\vec{B}_{\rm eff }
\propto \vec{n} \times \nabla \vec{n}$ consequently is
perpendicular to both the magnetization and its gradient. It is
this current-induced spin polarization that contributes to the
effective field for the magnetization and leads to current-induced
torques. The above argument holds in an analogous form for
ferromagnets and confirms that the magnetization-direction
dependence of $\langle \vec{s}_{\perp} \rangle$ for ferromagnets,
is the same as its N\'e\`el vector dependence for
antiferromagnets. Note that at the level of the effective
hamiltonian in Eq.~(\ref{eq:SOHamiltonain}) the main difference
between antiferromagnetic and ferromagnetic case is the difference
in $H_0$, which in the ferromagnetic case contains a constant
exchange splitting and in the antiferromagnetic case an
alternating one. Current-induced spin polarization has been
studied in paramagnetic semiconductors
\cite{Sih2005,Aronov1989,Edelstein1990}, which shows that they do
not require a net nonzero exchange splitting, which, in turn,
explains why the current-induced torques in antiferromagnets are
generally nonzero.

\noindent {\it Discussion \& Conclusions} --- So far we have
ignored conduction-electron spin relaxation which in the
ferromagnetic case is known to result in an additional
contribution $-\beta\vec{\nabla}\vec{\Omega}$ to the spin density
 that is parameterized by the
dimensionless constant $\beta$ \cite{zhang2004,barnes2005}. As we
have shown the spin density has, without spin relaxation, the same
form for both ferromagnetic and antiferromagnetic textures. It is
therefore reasonable to assume that spin relaxation will lead to a
similar correction for antiferromagnets. Hence, we expect that the
spin density takes the form $
\left<s_{\perp}\right>\propto\vec{n}\times\left(\vec{\upsilon}\cdot\vec{\nabla}\right)\vec{n}+\beta_{\rm
afm} \left(\vec{\upsilon}\cdot\vec{\nabla}\right)\vec{n}$ when we
include conduction-electron spin relaxation that leads to the
correction phenomenologically parameterized by the dimensionless
constant $\beta_{\rm afm}$. This will change the equation of
motion Eq.(\ref{eq:AFMeqm}) which now becomes
\begin{equation}
\label{spinrelaxEqm} \frac{\partial^{2}\vec{n}}{\partial
t^{2}}+\frac{\partial}{\partial
t}\left([\vec{\upsilon}\cdot\vec{\nabla}]\vec{n}\right)-c^{2}\nabla^{2}\vec{n}+\beta_{\rm
afm }\frac{\partial}{\partial
t}\left(\vec{n}\times[\vec{\upsilon}\cdot\vec{\nabla}]\vec{n}\right)=0.
\end{equation}
In future work we intend to explore the consequences of this
equation of motion for the current-driven motion of
antiferromagnetic domain walls, taking into account also
anisotropy and magnetization relaxation. Further studies will also
include exploring other aspects of the effective spin-orbit
coupling discussed above.

In conclusion, we have derived a general equation of motion for
the influence of an electric current on the long-wavelength
dynamics of a smooth antiferromagnetic magnetization texture.
Although we have only presented numerical calculations for a
simple model, we believe that the form of this equation of motion
is quite general, because the arguments leading to
Eqs.~(\ref{eq:SOHamiltonain})~and~(\ref{spinrelaxEqm}) do not
depend on details.

This work was supported by the Stichting voor Fundamenteel
Onderzoek der Materie (FOM), the Netherlands Organization for
Scientific Research (NWO), and by the European Research Council
(ERC) under the Seventh Framework Program (FP7).

\end{document}